\documentclass[runningheads]{llncs}

\usepackage[inline]{enumitem}
\usepackage{graphicx}

\usepackage{xcolor}

\begin{document}

\title{POSTER: Towards Secure 5G Infrastructures for Production Systems}

\author{%
Martin Henze\inst{1,2} \and
Maximilian Ortmann\inst{3} \and
Thomas Vogt\inst{1} \and \\
Osman Ugus\inst{4} \and
Kai Hermann\inst{5} \and
Svenja Nohr\inst{6} \and
Zeren Lu\inst{4} \and \\
Sotiris Michaelides\inst{1} \and
Angela Massonet\inst{3} \and
Robert H. Schmitt\inst{1,3}
}

\authorrunning{M.\ Henze et al.}

\institute{%
$^1$ RWTH Aachen University,
$^2$ Fraunhofer FKIE,
$^3$ Fraunhofer IPT, \\
$^4$ Swissbit Germany AG,
$^5$ Utimaco IS GmbH, and
$^6$ oculavis GmbH
\\Corresponding author: \email{henze@spice.rwth-aachen.de}
}

\maketitle

\begin{abstract}
To meet the requirements of modern production, industrial communication increasingly shifts from wired fieldbus to wireless 5G communication.
Besides tremendous benefits, this shift introduces severe novel risks, ranging from limited reliability over new security vulnerabilities to a lack of accountability.
To address these risks, we present approaches to 
\begin{enumerate*}[label=(\roman*)]
\item prevent attacks through authentication and redundant communication,
\item detect anomalies and jamming, and
\item respond to detected attacks through device exclusion and accountability measures.
\end{enumerate*}
\end{abstract}

\begin{keywords}
5G \and Network Security \and Industrial Networks \and Wireless Networks
\end{keywords}

\section{Introduction}

The digital transformation of production requires a change from conventional to flexible and networked systems such as the Industrial Internet of Things (IIoT)~\cite{kehl2020comparison,serror2021challenges,dahlmanns2022missed}. 
5G communication offers the opportunity to advance the networking of production and thus plays a key role in the digitization of production. 
Yet, deploying 5G as the communication technology in industrial settings introduces new cybersecurity risks for manufacturing companies.

\begin{figure}[t]
\centering
\includegraphics{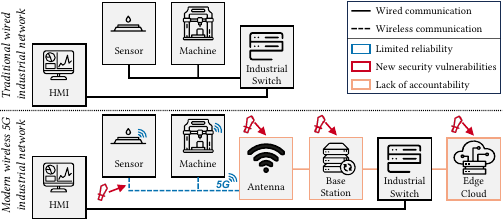}
\vspace{-1em}
\caption{Moving from wired to wireless industrial communication introduces novel risks: \textcolor[HTML]{0571B0}{limited reliability}, \textcolor[HTML]{CA0020}{new security vulnerabilities}, and \textcolor[HTML]{F4A582}{lack of accountability}.}
\vspace{-1em}
\label{fig:risks}
\end{figure}

These risks mainly result from the introduction of wireless communication into previously wired industrial networks as illustrated in Fig.~\ref{fig:risks}~\cite{bodenhausen2023challenges}.
The first concern involves the \emph{limited reliability of systems}. Current production systems are designed with high-reliable fieldbuses, optimized for specific use cases, posing a challenge to replicate in wireless 5G environments.
Second, by breaking up local fieldbus systems through the introduction of 5G, \emph{new security vulnerabilities} are emerging on the safety-critical production shop floor.
Finally, the integration of 5G systems into production increases the number of companies involved, leading to a \emph{lack of accountability}. %
Consequently, to capitalize on the promisingly significant benefits of 5G and wireless communication, it is requires to address the risks, especially w.r.t.\ security.

\textbf{Related Work.}
Several works study the challenges resulting from switching to 5G communication in general (not specific to production) \cite{Wen.2022,ahmad20175g}. 
Similarly, separate research focuses on securing industrial and IIoT networks without considering 5G \cite{serror2021challenges,wolsing2022ipal,dahlmanns2022missed}.
However, a systematic and holistic approach to securely deploy 5G  in real-world production systems is missing.

\textbf{Contributions.}
We report on our ongoing work to secure industrial 5G communication.
Our overarching goal is to design, implement, and validate a comprehensive toolbox of solutions for the prevention, detection, and response to the risks associated with using 5G for industrial communication.

\section{5G in Industrial Communication}
\label{sec:background}

Wireless communication is a key component of future smart factories, facilitating comprehensive automation, optimization, and flexibility through the scalable, networked operation of sensors and actuators, the integration of large computing resources, and the incorporation of IIoT devices on the shop floor.

Operating Technologies (OT) in manufacturing require communication technology to ensure reliability, availability, and real-time capability. 5G non-public networks are gaining attention for meeting these demands, offering superior performance and security compared to WiFi, Bluetooth, or LTE \cite{Caro.2022}. With capabilities like Massive Machine Type Communication (mMTC), and Ultra-Reliable Low-Latency Communication (URLLC), 5G addresses diverse scenarios, including IIoT needs for low latency and reliable communication.

However, the transition from wired, vendor-specific industrial solutions to multi-vendor wireless networks introduces new risks, such as insufficient integrability into existing IT infrastructures, new security vulnerabilities, and a lack of accountability.
Therefore, a widespread adoption of 5G in production is challenging due to security and legal related uncertainties in such diverse architectures.

\section{Secure 5G Infrastructures for Production Systems}

\begin{figure}[t]
\centering
\includegraphics[width=\textwidth]{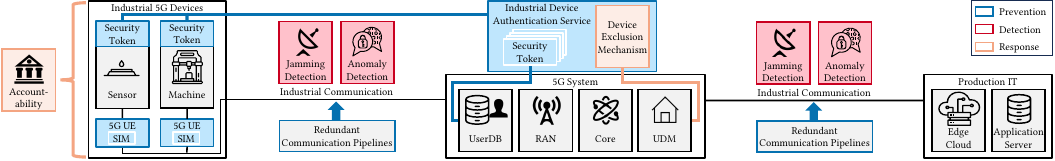}
\vspace{-1.5em}
\caption{Our secure 5G infrastructure for production systems provides mechanisms to \textcolor[HTML]{0571B0}{prevent}, \textcolor[HTML]{CA0020}{detect}, and \textcolor[HTML]{F4A582}{respond} to attacks.}
\vspace{-1.5em}
\label{fig:overview}
\end{figure}

To realize secure 5G infrastructures for production systems, we propose a comprehensive approach (Fig.~\ref{fig:overview}) encompassing %
\begin{enumerate*}[label=(\roman*)]
\item authentication, authenticity, and redundancy methods to \emph{prevent} attacks,
\item anomaly and jamming \emph{detection}, and
\item device exclusion and accountability measures to \emph{respond} to attacks.
\end{enumerate*}

\break

\subsection{Prevention}\label{sec:approach:prevention}
To secure industrial devices in 5G networks, we provide measures to mitigate threats such as impersonation, physical compromise, and machine-in-the-middle.

\textbf{Industrial Device Authentication.}
To mutually authenticate industrial devices, we develop a hardware security token-based industrial device authentication (IDA) service, independent of the network, with a 5G system interface for device exclusion (cf.\ Sec.~\ref{sec:approach:response}).
The IDA service periodically authenticates all industrial 5G devices using a challenge-response-based mutual entity authentication protocol. 
A hardware security token stores cryptographic material, and keys are provided by a certificate authority through the X.509 PKI standard. Certificates can be requested and revoked via REST, EST, CMP, SCEP, or ACME.
The validity status of each certificate is saved at an OCSP responder and verified via OCSP \cite{rfc6960} or a revocation list. 
Furthermore, our key establishment method based on the IDA service and secure tokens addresses threats that cannot be solved by device authentication alone, e.g., by providing keys for security protocols supporting pre-shared keys, such as TLS.
By combining identifiers, i.e., device identity and SIM, with mutual authentication between IDA service and devices, unauthenticated devices can be detected and excluded. %

\textbf{Redundant Communication.}
To enable industrial real-time applications, we develop a deterministic 5G Time-Sensitive Networking (TSN) infrastructure with redundant, hardware-separated communication pipelines.  As TSN operates at the OSI MAC layer (Layer 2), higher-level security mechanisms offer no protection to TSN data~\cite{macLayerSecWatson}.
Hence, we must implement security at Layer 2 or below to mitigate TSN-specific security risks, and satisfy low latency and high performance requirements.
MACsec~\cite{Macsec} safeguards authenticity and optionally confidentiality in TSN-based communication. 
We obtain MACsec keys from the key establishment mechanism of the IDA service to transfer the trust and security gained from device authentication into a TSN-based redundant communication.
\subsection{Detection}\label{sec:approach:detection}
We propose complementary approaches to detect any remaining anomalies in industrial 5G communication as well as jamming attacks on the physical layer.

\textbf{Anomaly Detection.}
Unlike intrusion detection in office networks and data centers, anomaly detection in industrial networks leverages the predictability and determinism of industrial communication.
It is proven effective in identifying subtle anomalies in timing, order of communication, or recognizing invalid physical states \cite{wolsing2022ipal,aoudi2018truth,choi2018detecting}. 
However, when considering wireless communication, unique challenges prevent the adoption of existing anomaly detection for industrial networks. 
Most notably, inherent wireless characteristics such as the use of a shared medium result in less predictable communication patterns. %
Likewise, the use of wireless communication further demands for encrypted communication, preventing detection approaches relying on deep packet inspection \cite{wolsing2022ipal}.

To address these issues, we develop a method for detecting anomalies in industrial 5G by incorporating wireless characteristics, such as packet drops and retransmissions, to existing timing- and sequence-based solutions. Concretely, we extend the IPAL framework for industrial intrusion detection \cite{wolsing2022ipal} to realize anomaly detection for less predictable wireless communication.
To optionally also allow for the inclusion of process state information despite the use of encrypted communication, we further extend the IPAL framework to decrypt TLS.

\textbf{Jamming Detection.}
Denial of service attacks on the physical layer of 5G through radio jamming pose a major threat to highly connected production systems. 
Easily accessible jamming devices can be used to disrupt the connectivity of assets, resulting in machine downtime or, in the worst case, causing harm to humans.
To counter corresponding threats, we develop a jamming detection and localization framework using software-defined radios tested in a realistic shop floor environment.
The framework will enable the development of mitigation and response strategies tailored to industrial environments.
\subsection{Response}\label{sec:approach:response}
Supplementing prevention and detection, response mechanisms include technical approaches such as device exclusion as well as legal accountability guidelines.

\textbf{Device Exclusion.}
Compared to information technology, OT systems prioritize aspects such as performance, resource limitations, and availability in relation to potential system failures. 
Therefore, two device exclusion mechanisms will be examined:
\begin{enumerate*}[label=(\roman*)]
\item a PKI-based devices exclusion mechanisms executed by the IDA service when the authenticity of an industrial device cannot be verified and 
\item a 5G-based device exclusion approach that blocks SIM cards in case of detected malicious activities. 
\end{enumerate*}
As PKI-based response, we can suspend and/or revoke the certificate of the malicious industrial device using corresponding PKI APIs.
However, as the industrial device still remains integrated into the 5G network, this strategy requires entities to validate certificates via the OCSP responder for secure certificate-based communication whenever connecting with an industrial device.
To lift this requirement, a 5G-based response allows to block the SIM responsible for linking the malicious industrial device to the 5G network, e.g., by utilizing the 5G API to remove the corresponding subscriber from the 5G core network. The device's credentials, such as the SUPI, are made available to the IDA service for this purpose.

\textbf{Accountability.} To address legal questions of responsibility and accountability for production failures due to vulnerabilities in 5G communication infrastructures, multi-vendor architectures, and networked production facilities, a legal assessment of the developed approaches complements our ongoing work. This includes considerations for legal requirements in cybersecurity, the strengthened powers of supervisory authorities, and questions regarding the use of communication data under civil and data protection law. Furthermore, the impact of planned regulations such as the European NIS2 Directive, the European Data Act, and the European Cyber Resilience Act is also covered.

\vspace{-0.1mm}
\section{Outlook \& Conclusion}

Our collaborative effort between industry and academia aims to comprehensively address cyberattacks on 5G communication in industrial production by developing, implementing, and integrating preventive, detection, and response measures. We evaluate these approaches in industry-driven use cases, focusing on machine monitoring and remote maintenance within the 5G-Industry Campus Europe~\cite{Koenig_Kehl_2022}, covering a machine hall and 1 km\textsuperscript{2} of outdoor space.
This enables us to research and develop 5G technologies for complex industrial applications and validate them in practical settings to ensure that our solutions are practical and adaptable to new use cases and environments.

\textbf{Acknowledgments.}
This work has been funded by the German Federal Office for Information Security (BSI) under project funding reference numbers 01MO23016A, 01MO23016B, 01MO23016C, 01MO23016D, and 01MO23016G (5G-Sierra).
The authors are responsible for the content of this publication.

\end{document}